\documentclass{elsart}

\input epsf.sty
\usepackage[centertags]{amsmath}
\usepackage{amsfonts}
\usepackage{amssymb}
\newcommand{\R}{r}
\newcommand{\Nb}{\vec{N}'}
\newcommand{\N}{\vec{N}}
\newcommand{\n}{\vec{n}}

\newcommand{\NC}{\vec{{\mathcal N}}}
\newcommand{\D}{\Delta}

\newcommand{\rr}{\vec{r}}

\def\mref#1{(\ref{#1})}
\def\eqref#1{(\ref{#1})}

\begin{document}

\begin{frontmatter}

\title{Bianchi surfaces. Integrability in arbitrary parametrization}
\author[l,w]{
Maciej Nieszporski
}
\author[w]{Antoni Sym
}
\maketitle

\begin{abstract}
We discuss  integrability of normal field equations of arbitrary parametrised Bianchi surfaces.
A novel geometric definition of Bianchi surfaces is presented as well as B\"acklund
transformation for the normal field equations in arbitrary chosen surface parametrization.
\end{abstract}
\address[l]{Department of Applied Mathematics, University of Leeds, Leeds LS2 9JT, UK}
\address[w]{Katedra Metod Matematycznych Fizyki,
Uniwersytet Warszawski
ul. Ho\.za 74, 00-682 Warszawa, Poland}

\end{frontmatter}

\section{Introduction}
In the second half of XIX century Luigi Bianchi had paid attention to
transformations of pseudo-spherical surfaces 
in Euclidean space ${\mathbb E}^3$ \cite{BiaPS}.
The transformations had been generalised by B\"acklund  \cite{Bac}.
It was then the theory of integrable nonlinear systems began.
In the case of the pseudo-spherical surfaces
most of the considerations were carried out in asymptotic coordinates and led to transformations of
\begin{itemize}
\item
Gauss-Mainardi-Codazzi  equations that reduce in the case to single equation
\[\phi,_{uv}=\sin \phi \]
and B\"acklund transformation for the  sine-Gordon equation
appeared in this context first time \cite{Dar}
\item
unit normal field (of the surface) equation
\[\n,_{uv}= f \n
\quad
\n \cdot \n= 1\]
nowadays termed (hyperbolic) nonlinear  $\sigma$-model \cite{Poh}.
Vectors as usual denotes functions that take values in the tangent bundle of 3D affine space. The affine space we enrich with (pseudo) scalar product
$\vec{A} \cdot \vec{B}:=A_1 B_1+\varepsilon(A_2 B_2+A_3 B_3)$
making the space  either Euclidean one ${\mathbb E}^3$ ($\varepsilon=1$)
or Minkowski one ${\mathbb M}^3$ ($\varepsilon=-1$) 
\end{itemize}
The case of spherical surfaces governed by sinh-Gordon equation 
(or elliptic nonlinear $\sigma$-model) was treated separately
in isothermally-asymptotic coordinates \cite{BiaSS}.
It was Bianchi who generalised the results on pseudo-spherical surfaces to surfaces
that Gauss curvature $K$ in asymptotic coordinates is of the form
\[K=-\frac{1}{[U(u)+V(v)]^2} \]
and in consequence B\"acklund transformation  for the system
\begin{equation}
\label{BS}
\N,_{uv}= f \N
\quad
\N \cdot \N= U(u)+V(v)
\end{equation}
appeared \cite{BiaB,Bia05}.  
There is a tendency to call 
underlying surfaces Bianchi surfaces since they were "resurrected" 
in the papers \cite{Cen,LeSy}, while
the system of equations \mref{BS} 
or corresponding Gauss-Mainardi-Codazzi  equations (c.f. \cite{LeSy})
\begin{equation}
\label{AS}
\begin{array}{c}
\psi,_{uv}+(\frac{r,_v s \sin{\psi}}{2 r t}),_v+
(\frac{r,_u t \sin{\psi}}{2 r s}),_u-ts \sin{\psi}=0\\
s,_v=\frac{1}{2 r} ( t \cos{\psi} r,_u-s r,_v )\\
t,_u=\frac{1}{2 r} ( s \cos{\psi} r,_v-t r,_u )\\
r=U(u)+V(v)
\end{array}
\end{equation}
are both referred to as Bianchi system. 

This is only a part of the story.
The point is that the nonlinear system \mref{BS}
has its celebrated place
in the contemporary physics literature regardless of Bianchi results.
We have  already referred to the nonlinear $\sigma$-model \cite{Poh}
(or rather their non-isospectral extension see e.g. \cite{LeSy}) often called chiral model 
or harmonic map (onto a sphere)
\cite{Mis},
but also hyperbolic version of the Ernst  equation of general relativity 
(see \cite{Ern} for elliptic Ernst equation and
\cite{Nut} for hyperbolic one) 
\begin{equation}
(\varepsilon +\xi\bar{\xi})(\xi,_{uv}+\frac{r,_v}{2r}\xi,_u+\frac{r,_u}{2r}\xi,_v)=2 \bar{\xi} \xi,_u \xi,_v
\end{equation}
arises (in case $\varepsilon=-1$) when we normalise the vector field  $\N$ obeying  \mref{BS} first
\[\n=\frac{\N}{\sqrt{r}} \quad r:=U(u)+V(v)\]
and then make stereographic projection of the unit vector field $\n$ 
onto the complex plane
$\xi:=\frac{n_2+i n_3}{1+n_1}$.

What strikes in above introduction is that although  the objects
under considerations  are purely geometric (surfaces, rectilinear congruences)
and most of constructions is of geometric nature (see e.g. Finikov's monograph \cite{Fin}),
whenever integrable phenomena (B\"acklund transformations, permutability theorems etc.) are discussed the authors always confine themselves to particular parametrization
 of the  surface. 
Both at the beginning of the XX century \cite{Byu} and
in modern theory of integrable systems \cite{Taf} geometric characterisation of Bianchi
surfaces was known 
but transformations of the surfaces was carried on 
either in asymptotic parametrization or in isothermally-asymptotic  parametrization of the surface.

Moreover the contemporary theory of integrable systems seems to make
"para\-me\-trisation addiction" even stronger. 
We are able to indicate only two papers in the contemporary literature that 
discuss Darboux-B\"acklund
transformations of surfaces without referring to a particular parametrization.
The first one is about Darboux-B\"acklund
transformations for isothermic surfaces \cite{Jer}.
The second one discuss
integrability of GMC equations of Bianchi surfaces \cite{Jan}. Here we concentrate on the normal field equations of Bianchi surfaces instead. 
The "addiction" to surface parametrization is extremely visible in the so called difference (integrable) geometry
\cite{Sau,BoPi,BoSu}
where discretizations of particular  nets has been considered so far.

In our opinion restricting to particular parametrization, 
even though it is convenient and mentioned asymptotic nets are  geometrical objects themselves,
is not satisfactory. The aim of the paper is to present B\"acklund transformation for
a normal field of Bianchi surfaces in arbitrary  parametrization.

The essence of the work can be explained
in a few sentences. Namely,
let a regular surface, with a non-zero Gauss curvature $K$  and  unit normal field $\n_0$,
is given. Define normal field
\begin{equation}
\label{NCi}
{\NC}:= \frac{1}{\sqrt[4]{|K|}} {\n_0}
\end{equation}
By $*$ we denote Hodge dualization with respect to second fundamental form of the surface.
Bianchi surface is a surface for which (note we are unifying hyperbolic and elliptic case)
\begin{eqnarray}
\label{BSGA}
*d*d{\NC} = f {\NC}
\\
\label{BSGB}
*d*d({\NC} \cdot {\NC})=0
\end{eqnarray}
holds, where $f$ is a scalar function treated as an additional dependent variable.
This point of view were prompted  by Tafel \cite{Taf}.
The novelty of the paper is introduction of
distinguished normal field $\NC$ and deriving B\"acklund transformations
in arbitrary chosen parametrization of the surface.

\section{Extended Moutard transformation}
Classical Moutard transformation \cite{Mou} can be extended so that to act on general self-adjoint second
order  differential equation in two independent variables \cite{NSD}. 
Namely, map $\psi \mapsto {\psi}'$ given by
\begin{equation}
\label{matrc}
\left[ 
\begin{matrix}
(\theta {\psi}'),_x \cr (\theta {\psi}'),_y
\end{matrix}
\right] = \theta^2
\left[ 
\begin{matrix}
c&b\cr
-a&-c
\end{matrix}
\right]
\left[ 
\begin{matrix}
\left(\frac{\psi}{\theta} \right),_x \cr \left(\frac{\psi}{\theta} \right),_y
\end{matrix}
\right]
\end{equation}
is the map from solution space of equation 
\begin{equation}
\label{f}
\begin{array}{c}
{\mathcal L}^f \psi =0 \\ \\
{\mathcal L}^f :=
a \partial^2_x + b \partial^2_y+ 2 c \partial_x \partial_y+
(a,_x+c,_y)
\partial_x+(b,_y+c,_x)\partial_y-f 
\end{array}
\end{equation}
to solution space of the equation
\begin{equation}
\label{LB}
\begin{array}{c}
{\mathcal L}' {\psi}' =0 \\
{\mathcal L}':={a}' \partial_x^2 + {b}' \partial_y^2 +2 {c}'\partial_x\partial_y
+({a}',_x+{c}',_y)\partial_x +({c}',_x+{b}',_y)\partial_y - {f}'
\end{array}
\end{equation}
provided that $\theta$ is a solution of eq. \mref{f} (we assume that functions $a$, $b$, $c$
are of class 
${\mathcal C}^1$
and both function $\D$  given by 
\begin{equation}
\label{Del}
\D:=ab-c^2
\end{equation}
and $\theta$
obeys condition $\forall (x,y) \in {\mathcal D}; \D \ne 0; \theta \ne 0$ as well)
\[ (a \theta,_x+c \theta , _y),_x+(b \theta,_y+c\theta,_x)=f \theta\]
The coefficients of (\ref{LB}) are related to coefficients of (\ref{f}) by
\begin{equation}
\label{bar}
\begin{array}{c}
{a}'=\frac{-a}{\D}, \qquad {b}'=\frac{-b}{\D},\qquad {c}'=\frac{-c}{\D},
\\
{f}'=\{
-[\frac{a}{\D} \frac{1}{\theta},_{x}+\frac{c}{\D} \frac{1}{\theta},_{y}]_x
-[\frac{b}{\D} \frac{1}{\theta},_{y}+\frac{c}{\D} \frac{1}{\theta},_{x}]_y
\} \theta
\end{array}
\end{equation}
An elementary  observation is
\begin{equation}
\label{det}
a'b'-c'^2=\frac{1}{ab-c^2}
\end{equation}
So we have
\[\D'= \frac{1}{\D}\]
and as a result
\begin{equation}
\label{D}
 \frac{1}{\sqrt{|\D '|}}(a',b',c')=- \frac{1}{\sqrt{|\D|}}(a,b,c)
\end{equation}
\section{Lelieuvre formulae}
We are considering ${\mathcal C}^2$ vector valued function 
$ \N : {\mathbb R}^2 \supset {\mathcal D} \to T_p{\mathbb E}^3 (T_p{\mathbb M}^3)$ that obeys
\begin{equation}
\label{eMv}
 (a \N,_x+c \N , _y),_x+(b \N,_y+c\N,_x)=f \N 
\end{equation}
Cross multiplication by $\N$ yields
\[ [(a \N,_x+c \N,_y)\times \N],_x+[(b \N,_y+c \N,_x)\times \N],_y =0\]
so there exists a potential $\rr$ such that
\begin{equation}
\label{Lel}
\begin{array}{c}
\rr,_x = (b \N,_y+c \N,_x) \times \N
\\
\rr,_y = \N \times (a \N,_x+c \N,_y)
\end{array}
\end{equation}
We interpret the potential $\rr$ as position vector of a surface. Then $\N$ is a vector field
normal to the surface, Gauss curvature is given by
\begin{equation}
\label{K}
K=\frac{1}{ (\N \cdot \N)^2(ab-c^2)}
\end{equation}
and the second fundamental form is
\begin{equation}
\label{fund}
II=\frac{1}{\sqrt{\D K}} Vol({\n_0};{\n_0},_x;{\n_0},_y)
(b dx^2+a dy^2-2c dxdy)
\end{equation}
where $\n_0$ denotes unit normal field and $Vol$ is the volume form of the space.
\section{Bianchi surfaces from extended Moutard transformation}
In this section we derive basic formulae defining Bianchi surfaces.
We apply Moutard transformation to vector valued function 
$ \N $ defined in previous section and obeying the self-adjoint equation
\begin{equation}
\label{N}
 (a \N,_x+c \N , _y),_x+(b \N,_y+c\N,_x)=f \N 
\end{equation}
We get
\begin{equation}
\label{MTN}
\begin{array}{c}
(\theta \Nb),_x=
\theta^2 [c\left(\frac{\N}{\theta}\right),_x+b\left(\frac{\N}{\theta}\right),_y]
\\
(\theta \Nb),_y=
-\theta^2[a\left(\frac{\N}{\theta}\right),_x+c\left(\frac{\N}{\theta}\right),_y]
\end{array}
\end{equation}
We define quantities 
\begin{equation}
\label{def}
p:=\N \cdot \Nb, \quad r:=\N \cdot \N \quad {r}':= \Nb \cdot \Nb 
\end{equation}
so in our notation Gauss curvature \mref{K} (see definition \mref{Del}) is
\begin{equation}
\label{Kn}
K=\frac{1}{r^2 \D} 
\end{equation}
and without lost of generality we assume that $r>0$ in the domain ${\mathcal D}$.
From equations obtained by scalar multiplication of eqs. \mref{MTN} by $\N$ and $\Nb$
one can infer 
\begin{equation}
\label{conclcB}
\begin{array}{c}
\frac{1}{2} r',_x-\D\frac{1}{2}r,_x-c p,_x-b p,_y+\frac{\theta,_x}{\theta}(r'+r\D)=0
\\
\frac{1}{2} r',_y-\D\frac{1}{2}r,_y+a p,_x+c p,_y+\frac{\theta,_y}{\theta}(r'+r\D)=0
\end{array}
\end{equation}
Equation defining Bianchi rectilinear congruences (for Bianchi rectilinear congruences 
 see \cite{Fin,NiSy} and references therein) is
\[r'+r \D =0\]
or in virtue of \mref{det}
\begin{equation}
\label{defcB}
r' \sqrt{|\D '|}+\epsilon r \sqrt{|\D|}=0
\end{equation}
i.e. (see eq. \mref{Kn})
\begin{equation}
\label{condition}
K'+ \epsilon K=0
\end{equation}
where $\epsilon:=\hbox{sgn} (ab-c^2)$
In presence of constrain \mref{defcB} formulae \mref{conclcB} take form
\begin{equation}
\label{conclcM}
\begin{array}{c}
\epsilon  (r\sqrt{| \D|}),_x+\frac{c}{\sqrt{|\D|}} p,_x+\frac{b}{\sqrt{|\D|}} p,_y=0
\\
\epsilon  (r\sqrt{|\D|}),_y-\frac{a}{\sqrt{|\D|}} p,_x-\frac{c}{\sqrt{|\D|}} p,_y=0
\end{array}
\end{equation}
On eliminating function $p$ and using $r\sqrt{\epsilon \D}=\frac{1}{\sqrt{\epsilon K}}$
we obtain equation
\begin{equation}
\label{beq}
\left[\frac{\partial}{\partial x} 
\left(\frac{a}{\sqrt{| \D |}}\frac{\partial}{\partial x}+
\frac{c}{\sqrt{|\D |}}\frac{\partial}{\partial y}\right) +
\frac{\partial}{\partial y} 
\left(\frac{c}{\sqrt{| \D |}}\frac{\partial}{\partial x}+
\frac{b}{\sqrt{| \D |}}\frac{\partial}{\partial y}\right) \right]
\frac{1}{\sqrt{| K |}}=0
\end{equation}
which characterises, together with eq. \mref{N}, Bianchi surfaces.
\section{Bianchi surfaces in parametrization free language}

One can rewrite the results of our considerations in parametrization independent language.
Let $\n_0$ denotes unit normal field. Define normal field
\[{\NC}:= \frac{1}{\sqrt[4]{|K|}} {\n_0}\]
where $K$ is Gauss curvature of our surface.
By $*$ we denote Hodge dualization with respect to second fundamental form \mref{fund}.
Equation \mref{beq} becomes
\[*d*d({\NC} \cdot {\NC})=0\]
while equation \mref{N} rewritten in terms of vector field ${\mathcal N}$ takes form
\[*d*d{\NC} = \tilde{f} {\NC}\]
where $\tilde{f}= f/\sqrt{|\D|}$.
We come to the proposition:
\begin{prop}
Bianchi surface is surface for which
\begin{equation}
\label{Bdef}
\begin{array}{c}
*d*d{\NC} = \tilde{f} {\NC}
\\
*d*d({\NC} \cdot {\NC})=0
\end{array}
\end{equation}
holds for the vector normal field
\[{\NC}:= \frac{1}{\sqrt[4]{|K|}} {\n_0}\]
where $K$ is Gauss curvature of the surface and $\n_0$ is unit field normal
to the surface.
\end{prop}
\section{Orthonormal frame and rotation coefficients}
We associate with the surface orthonormal frame
\[ (\n_0,\n_1,\n_2)\]
where $\n_0$ is unit vector field normal to the surface so
\[\n _0:= \frac{\N}{\sqrt{\R}}\]
We confine ourselves in the paper to the case $\n_0 \cdot \n_0=1$,
$\n_1 \cdot \n_1=\varepsilon$ ,$\n_2 \cdot \n_2=\varepsilon$ in the domain (so in the case of Minkowski
space we assume the normal vector is spatial one).

The motion of the frame is described by formulae
\begin{equation}
\label{mf}
\begin{array}{c}
\n_A,_x ={p_A}^B \n_B \quad \n_A,_y ={q_A}^B \n_B
\end{array}
\end{equation}
Since the reper is orthonormal matrices ${p_A}^B$ and ${q_A}^B$ are  
either $so(3)$ ($\varepsilon=1$) or $so(1,2)$ ($\varepsilon=-1$) valued,
they are called rotation coefficients.

Compatibility conditions of system \mref{mf} read
\begin{equation}
\label{rccc}
\begin{array}{c}
({p_A}^C),_y+{p_A}^B{q_B}^C=({q_A}^C),_x+{q_A}^B{p_B}^C
\end{array}
\end{equation}
Rotation coefficients obey also equations
\begin{equation}
\label{rcno}
\begin{array}{l}
\, [\R(a{p_0}^{\nu}+c{q_0}^{\nu})],_x+[\R(c{p_0}^{\nu}+b{q_0}^{\nu})],_y+ \\
\hfill \R[a{p_0}^{B}{p_B}^{\nu}+b{q_0}^{B}{q_B}^{\nu}+
c({p_0}^{B}{q_B}^{\nu}+{q_0}^{B}{p_B}^{\nu})]=0\\
\, [(a(\sqrt{\R}),_x+c(\sqrt{\R}),_y),_x+
(c(\sqrt{\R}),_x+b(\sqrt{\R}),_y),_y] / \sqrt{\R}+    \\
\hfill a{p_0}^{B}{p_B}^{0}+b{q_0}^{B}{q_B}^{0}+
c({p_0}^{B}{q_B}^{0}+{q_0}^{B}{p_B}^{0})=f 
\end{array}
\end{equation}
as a consequence of the fact that field $\n_0$ is proportional to $\N$ that satisfies eq. \mref{N}.

\section{B\"acklund transformation}
We decompose the $ \N '$ in the orthonormal basis we described in the previous section
\begin{equation}
\label{defNp}
\begin{array}{l} 
\theta \N '= x^A \n_A 
\end{array}
\end{equation}
and substitute in the extended Moutard transformation
(we use convention that Greek indexes $\mu$ and $\nu$ goes from $1$ to $2$ while
capital Latin from $0$ to $2$)
\begin{equation}
\label{matrc1}
\left[ 
\begin{matrix}
(\theta \N '),_x \cr (\theta \N '),_y
\end{matrix}
\right] = \theta^2
\left[ 
\begin{matrix}
c&b\cr
-a&-c
\end{matrix}
\right]
\left[ 
\begin{matrix}
\left(\frac{\N}{\theta} \right),_x \cr \left(\frac{\N}{\theta} \right),_y
\end{matrix}
\right]
\end{equation}
Taking into account that due to \mref{defNp} function $\theta$ can be given in terms
of functions $x^0$, $r$ and $p$, namely
\[\theta= x^0\frac{\sqrt{r}}{p} \] 
we obtain that coefficients $x^A$ satisfies linear system
\begin{equation}
\label{ls}
\begin{array}{l}
x^0,_x=\frac{{p}^2}{{p}^2+\R^2 \D} \left\{
[(\frac{\R}{p}c-1){p_{\mu}}^0+\frac{\R}{p} b {q_{\mu}}^0] x^{\mu}+
\frac{\R}{p^2}[(c+ \frac{\R}{p}\D)p,_x+b p,_y]x^0 \right\}
\\
x^{\mu},_x=-x^A{p_A}^{\mu}+\frac{\R}{p}(c{p_0}^{\mu}+b{q_0}^{\mu}) x^0\\
x^0,_y=-\frac{{p}^2}{{p}^2+\R^2 \D} \left\{
[(\frac{\R}{p}c+1){q_{\mu}}^0+\frac{\R}{p} a {p_{\mu}}^0] x^{\mu}+
\frac{\R}{p^2}[(c- \frac{\R}{p}\D)p,_y+a p,_x]x^0 \right\}
\\
x^{\mu},_y=-x^A{q_A}^{\mu}-\frac{\R}{p}(a{p_0}^{\mu}+c{q_0}^{\mu}) x^0
\end{array}
\end{equation}
Since the rotation coefficients matrices are $so(3)$ ($so(1,2)$) valued compatibility conditions of the above linear system consist of \mref{rccc}, \mref{rcno} and two conditions
\begin{equation}
\begin{array}{l}
{p_{\mu}}^0 [ap,_x+cp,_y-r,_y \D- \frac{1}{2} {r} \D,_y+\\
\frac{r}{p}(ar,_x\D+cr,_y \D+\frac{1}{2} a {r} \D,_x+\frac{1}{2} c {r} \D,_y+p,_y \D)]+\\
{q_{\mu}}^0 [cp,_x+bp,_y+r,_x \D+ \frac{1}{2} {r} \D,_x+\\
\frac{r}{p}(br,_y\D+cr,_x \D+\frac{1}{2} b {r} \D,_y+\frac{1}{2} c {r} \D,_x-p,_x \D)]
=0
\\
\\
\left\{\frac{r}{p^2+r^2 \D}[ap,_x+(c-\frac{r}{p}\D)p,_y ]\right\},_x+
\left\{\frac{r}{p^2+r^2 \D}[(c+\frac{r}{p}\D)p,_x+bp,_y]\right\},_y=0
\end{array}
\end{equation}
that are satisfied due to \mref{conclcM}.
In addition in virtue of definitions \mref{def} and \mref{defNp} coefficients $x^A$ are subjected to the constraint
\begin{equation}
\label{cons}
\begin{array}{l}
(x^0)^2 (1+\frac{\R^2}{p^2} \D)+\varepsilon \left[(x^1)^2+(x^2)^2\right]=0
\end{array}
\end{equation}
Fortunately enough the quantity $(x^0)^2 (1+\frac{\R^2}{p^2} \D)+(x^1)^2+(x^2)^2$ is
a first integral of the linear system \mref{ls}. 
So one can choose constants of integration so that \mref{cons} holds.
Therefore we can formulate theorem
\begin{thm}[Transformations between normal fields of Bianchi surfaces]
\label{tw}
We assume that the position vector  $\vec{r}$ of a Bianchi surface is given so one can find
\begin{enumerate}
\item its Gauss curvature $K$,  
\item a normal field $\N$ to the surface  in particular unit field normal to the surface $\n_0$
and the normal field $\NC=\frac{1}{\sqrt[4]{|K|}} {\n_0}$,
we assume \[Vol({\n_0};{\n_0},_x;{\n_0},_y)\ne 0\]
\item functions $a$, $b$, and $c$ from the Lelieuvre formulae \mref{Lel}
\item an orthonormal frame $(\n_0,\n_1,\n_2)$ where 
$\n_1$ and $\n_2$ fields are tangent  to the surface
\item rotation coefficient $p_A^B$, $q_A^B$ through the formulae \mref{mf},
\item function $r$ given by $r:=\N \cdot \N=\frac{1}{\sqrt{K(ab-c^2)}}$ and then function $p$ 
integrating the system \mref{conclcM} (with a constant of integration say $k$).
\item solutions $(x^0,x^1,x^2)$ of the system \mref{ls} that obey constraint \mref{cons}
\item the normal field of new Bianchi surface (B\"acklund transform of the field $\N$)
\begin{equation}
\label{NP}
 \N ' := \frac{p}{\sqrt{r}} \frac{x^A}{x^0} \n_A 
\end{equation}
\item the position vector of the new surface
\begin{equation}
\label{rpr}
 \rr \, '= \rr + \N \times \N ' +\vec{c}
\end{equation}
\end{enumerate}
where $\vec{c}$ is a constant vector.
\end{thm}
Since the construction presented here is classical one \cite{Bia05}
we confine ourselves to outline of the proof
\begin{pf}

Ad 6

 The system \mref{conclcM} treated as a system on function $p$ is compatible due to the fact normal field $\N$ satisfies eq.
\mref{N} and \mref{beq} (constant of integration
of the system is spectral parameter!) and does define function $p$.

Ad 7

System \mref{ls} is compatible
due to the fact normal field $\N$ satisfies eq.
\mref{N} and \mref{beq}
and function $p$ is defined through \mref{conclcM}

Ad 8

Proof of the crucial eighth point  splits in two parts.
Firstly,  we define $\theta=x_0\frac{\sqrt{r}}{p}$ and we verify that $\N '$
given by \mref{NP} is related to $\N$ through a Moutard transformation \mref{MTN}.
It follows that  $\Nb$ solves \mref{N} with a function $f'$ and \mref{D} holds.
Secondly, since we imposed constraint \mref{cons} we get
\[r ' :=\Nb \cdot \Nb=(c^2-ab) r\] and \mref{K} as a result so $K'$ satisfies 
\mref{beq}.

Ad 9 We cross multiply formulae \mref{MTN} by $\N$ and by $\Nb$.
From the four equations obtained this way we can infer 
\[(\rr \, '- \rr - \N \times \N ' ),_x=0=(\rr \, '- \rr - \N \times \N ' ),_y\]
and therefore \mref{rpr} holds.



\quad
\end{pf}

We end paper with two comments.
First, we presented the transformation  acting on an arbitrary normal field of a Bianchi surface. To have an auto-B\"acklund transformation for a partial differential equation,
one has to confine himself to the distinguished field $\vec{\mathcal N}$.
In this case we receive B\"acklund transformation for the system

\begin{equation}
\label{e1}
\left[\frac{\partial}{\partial x} 
\left(A\frac{\partial}{\partial x}+
C \frac{\partial}{\partial y}\right) +
\frac{\partial}{\partial y} 
\left(C\frac{\partial}{\partial x}+
B \frac{\partial}{\partial y}\right) \right]
\NC =f \NC
\end{equation}
\begin{equation}
\label{e2}
\left[\frac{\partial}{\partial x} 
\left(A\frac{\partial}{\partial x}+
C \frac{\partial}{\partial y}\right) +
\frac{\partial}{\partial y} 
\left(C\frac{\partial}{\partial x}+
B \frac{\partial}{\partial y}\right) \right]
\NC \cdot \NC=0
\end{equation}
where $A$, $B$, $C$ are given by $(A,B,C)=\frac{1}{\sqrt{|ab-c^2|}}(a,b,c)$ so they obey constraint $AB-C^2=\pm 1$ and are conserved (up to irrelevant sign) under the transformation
(due to \mref{D}). 

\begin{cor}[B\"acklund transformation for the system \mref{e1}-\mref{e2}]~
\newline
The transformation given in the theorem \ref{tw} applied to the distinguished normal vector
field $\NC$ provide us with auto-B\"acklund transformation for the system \mref{e1}-\mref{e2}
where $\NC$ and $f$ are dependent variables of the system equations while
$A$, $B$, $C$ are given functions obeying constraint $AB-C^2=\pm 1$.
\end{cor}

Second,
in elliptic case starting from real valued $\NC$  one obtain   pure imaginary vector valued function  $\NC '$. To obtain real solution of the system \mref{Bdef} one has to apply
the transformation to $\NC '$ 
or alternatively make use of the permutability theorem (nonlinear superposition principle) we end the paper with
\begin{thm}[Permutability theorem]
Let a solution $\NC$ of the system of equations \mref{e1}-\mref{e2} is given
as well as  two B\"acklund transforms $\NC^{(1)}$ and $\NC^{(2)}$ of $\NC$
that correspond to constants of integration say $k_1$ and $k_2$ of the system \mref{conclcM}.
Applying the B\"acklund transformation to the solutions $\NC^{(1)}$ and $\NC^{(2)}$
and taking constants of integration of  \mref{conclcM} as $k_2$ and $k_1$ respectively
one can find solutions $\NC^{(12)}$ and $\NC^{(21)}$ such that   
$\NC^{(12)}=\NC^{(21)}$ and are given by
\begin{equation}
\NC^{(12)}=\epsilon 
\left[ -\NC +\frac{2 (\NC^{(1)}-\NC^{(2)})\cdot \NC}{(\NC^{(1)}-\NC^{(2)})\cdot(\NC^{(1)}-\NC^{(2)})}
(\NC^{(1)}-\NC^{(2)})\right]
\end{equation}
\end{thm}

{\bf Acknowledgements}
MN is grateful to Jan Cie\'sli\'nski and Frank Nijhoff for useful discussions. 
MN was supported at initial stage of the work
 by Polish KBN grant
1 P03B 017 28 while at the main stage (starting from 1st April 2005) 
solely by the European Community under a  Marie Curie Intra-European Fellowship,
 contract no MEIF-CT-2005-011228.  

\end{document}